\documentclass[pss]{wiley2sp} % provides pss two-column style
\usepackage{amsmath}

\begin{document}

\title{Quantification of local geometry and local symmetry in models of disordered materials}
\titlerunning{Local geometry from PDF}
\author{Matthew J. Cliffe and Andrew L. Goodwin\textsuperscript{\Ast}}
\authorrunning{Cliffe and Goodwin}
\mail{e-mail  \textsf{andrew.goodwin@chem.ox.ac.uk}, Phone: +44 (0)1865 272137, Fax: +44 (0)1865 272690}
\institute{Department of Chemistry, University of Oxford, Inorganic Chemistry Laboratory, South Parks Road, Oxford OX1 3QR, U.K.}
\received{XXXX, revised XXXX, accepted XXXX} % do not change, will be filled in by the publisher
\published{XXXX} % do not change, will be filled in by the publisher

\abstract{%
\abstcol{%
We suggest two metrics for assessing the quality of atomistic configurations of disordered materials, both of which are based on quantifying the orientational distribution of neighbours around each atom in the configuration. The first metric is that of geometric invariance: \emph{i.e.}, the extent to which the neighbour arrangements are as similar as possible for different atoms, allowing for variations in frame of reference. The second metric concerns the degree of local symmetry. We propose that for a set of configurations with equivalent pair correlations, ranking highly those configurations with low geometric
}{ invariance but with high local symmetry selects for structural simplicity in a way that does not rely on formal group theoretical language (and hence long-range periodic order). We show that these metrics rank a range of SiO$_2$ and \emph{a}-Si configurations in an intuitive manner, and are also significantly more sensitive to unphysical features of those configurations in a way that metrics based on pair correlations are not. We also report that implementation of the metrics within a reverse Monte Carlo algorithm gives rise to an energy landscape that is too coarse (at least in this initial implementation) for amorphous structure ``solution''.}}

%\titlefigure{fig0.png}
%\titlefigurecaption{%
%  This is the caption of the \emph{optional} abstract figure. If
%  there is no abstract figure here, the abstract text should be divided into both columns.}

\maketitle   % please do not remove

\section{Introduction}
In spite of the fact that the central axiom of crystallography---namely the existence of long-range structural periodicity---does not hold for disordered materials, diffraction methods remain a powerful tool in the study of the atomic-scale structure of glasses, liquids and other amorphous systems: total scattering measurements allow access to the experimental pair distribution function (PDF), which essentially represents a histogram of interatomic distances within a given material \cite{Egami2003}. Even in amorphous materials the PDF shows considerable structure at small distances, and a key question in the field concerns the extent to which these features are sufficient to produce a physically-sensible atomistic model of local structure for disordered systems \cite{Juhas2006,Cliffe2010,Juhas2010}. The principal motivations for asking (and attempting to answer) this question are twofold: first there is the enormous scientific value of understanding structure and its mechanistic or functional implications, and second there is the observation that many crucially-important materials of relevance to fields as diverse as biomaterials \cite{Weiner2005}, earth sciences \cite{Michel2007} and device physics \cite{Walsh2009,Sun2006} are either not available in a crystalline form or indeed gain their particular function only in the absence of crystallinity.

 \begin{figure*}%
\includegraphics*[width=\textwidth]{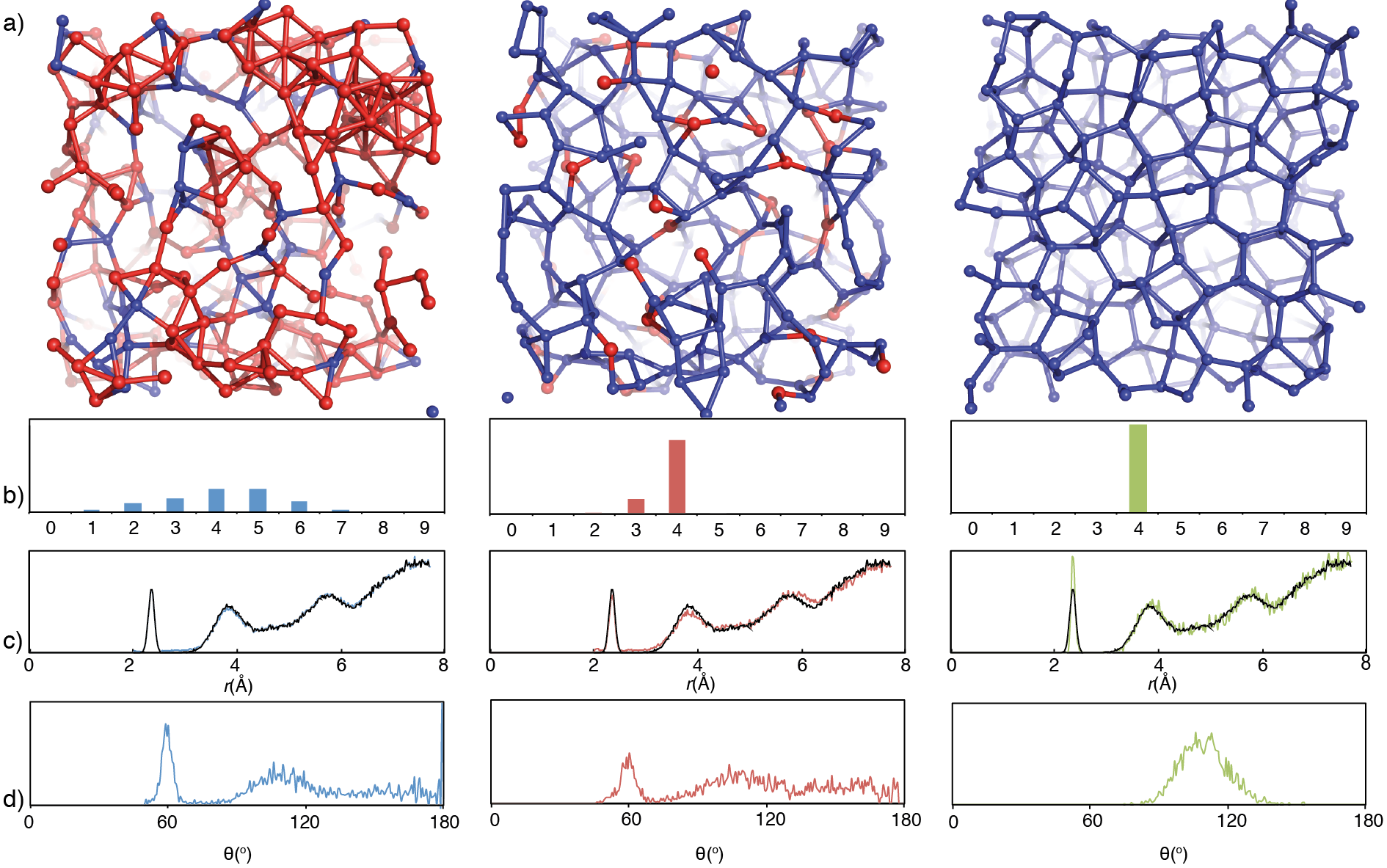}
\caption{A comparison of three models of \emph{a}-Si with similar pair correlations: (left) a configuration produced by RMC refinement against idealised \emph{a}-Si PDF data, (centre) a similar configuration but produced using the RMC+INVERT method \cite{Cliffe2010}, and (right) a model generated using the WWW algorithm \cite{Wooten1987,Wooten1985}. (a) Representations of portions of the configurations themselves, coloured according to coordination number (blue = four; red otherwise). (b--d) The corresponding (b) coordination number distributions, (c) PDFs, and (d) Si--Si--Si bond angle distribution functions. 
}
\label{fig1}
\end{figure*}

The process of translating the PDF into a three-dimensional structural model in many ways equates to the problem of extracting high-order correlation functions from the first and second-order correlations (to which diffraction data are sensitive, whether a material is long-ranged ordered or otherwise). That the PDF constrains the two-body correlations is clear: the nearest-neighbour distances can of course be read directly from the function itself. But a meaningful three-dimensional model requires at the very least an accurate depiction of bond and torsion angles, and the extent to which these higher-order correlations are constrained by the PDF has been questioned for some time \cite{Welberry1994,Evans1990}. In the limiting cases of long-range periodic order on the one hand, and vanishing two-body correlations on the other hand, the level of constraint is clear: the higher-order correlations are completely fixed in the former (we are used to the idea of using crystallography to determine bond angles and torsion angles in crystalline materials) and completely free in the latter (an excellent demonstration of this can be found in Ref.~\cite{Welberry1994}). Real disordered materials will lie between these two extremes, corresponding to the situation where a number of meaningfully-different three-dimensional models reflect the PDF equally well.

Amorphous silicon---perhaps the canonical disordered material by virtue of its chemical simplicity---illustrates this point clearly.  Its structure is widely believed to be well-described in terms of a continuous random network (CRN) of tetrahedrally-connected Si centres \cite{Biswas2004}. Arguably the highest quality atomistic models of \emph{a}-Si have been generated using the bond-switching algorithm of Wooten Winer and Weaire (WWW) \cite{Wooten1987,Wooten1985}, yielding CRNs that are consistent with a variety of experimental measurements and \emph{ab initio} calculations---X-ray or neutron diffraction, NMR \cite{Muller-Warmuth1982}, Raman spectroscopy \cite{Brodsky1978}, and band structure calculations \cite{Biswas2004}. One such model is shown on the right-hand side of Fig.~\ref{fig1}, together with the corresponding coordination number histogram, PDF and Si--Si--Si bond angle histogram. That the third-order (bond angle) correlations are well-defined is clear from the existence of a single narrow maximum in the bond angle-distribution function. So the absence of structural periodicity does not imply that higher-order correlations are poorly fixed. Yet if the PDF of \emph{a}-Si is used to drive atomistic refinement using a reverse Monte Carlo (RMC) algorithm, one obtains configurations with an almost perfect correspondence to the PDF but with very different higher-order correlations; one such configuration and the corresponding histogram functions are shown on the left-hand side of Fig.~\ref{fig1}. The bond angle correlation function is now significantly more complex, with the most structured contribution corresponding to a high concentration of Si$_3$ ``triangles'' \cite{Gereben1994}. These triangles are the most obvious incorrect feature of the structural model because they give rise to mid-gap electronic states that are not observed experimentally \cite{Biswas2004}.

Historically, the popular remedy for improving PDF-driven structural models is to incorporate in the fitting process one more additional constraints that act to improve the extent to which given correlation functions are physically sensible. So, for instance, constraints that encourage a single coordination number of four or that favour tetrahedral geometries or that penalise the formation of triangles are all capable of substantially improving RMC-type models of \emph{a}-Si \cite{Gereben1994}. Implicit in such an approach, however, is the \emph{a priori} assumption of a particular local structure: one decides before refinement that coordination numbers will equal four or coordination geometries should be tetrahedral, or that three-membered rings are not physical. While in the case of \emph{a}-Si these various assertions are no doubt true, we are nonetheless interested in the more difficult question of whether the PDF can be used to derive physically-sensible atomistic configurations in a model-independent approach. 

So the central idea we explore in this paper is that---in the absence of experimental data to suggest otherwise---the most physically-sensible solution amongst any set of structural models that reproduce the PDF equivalently well is likely to be that of greatest structural simplicity. This concept has its roots in the empirical tendency for real materials to have only a limited number of atomic environments---an observation reflected in Pauling's Rule of Parsimony \cite{Pauling1929}.\footnote{Originally stated as ``the number of essentially different kinds of constituents in a crystal tends to be small''.}

In a previous study \cite{Cliffe2010}, we introduced a similar concept---namely, that experimental knowledge of the number and distribution of local environments might be used to guide RMC refinement by minimising the variation in the individual atomic PDFs of relevant types of atoms. For the example of \emph{a}-Si, where it is known from NMR measurements that all Si atoms are in essentially the same local environment \cite{Shao1990}, this approach (dubbed ``INVERT'' = INVariant Environment Refinement Technique) had produced configurations with a much more sensible coordination number distribution, and also reduced the relative frequency of three-membered Si$_3$ triangles. A representative configuration obtained using an INVERT+RMC refinement is shown in the centre of Fig.~\ref{fig1}. Despite the noticeable improvements relative to an RMC-only refinement, the inadequacy of the INVERT+RMC model in terms of characterising higher-order correlations is also clear. This first crude implementation of the INVERT approach seeks only to minimise variations in the pairwise functions---the distances between an atom and its various shells of neighbours. Consequently we attribute the poor performance with respect to higher-order correlations to the fact that we are not applying INVERT in a manner which acts to encourage equivalent atoms to adopt equivalent spatial arrangements of their neighbours. So the INVERT+RMC configuration shown in Fig.~\ref{fig1} still represents a \emph{structurally-complex} solution to the PDF because it contains a large variety of different interatomic bond angles (and also higher-order correlations).

In this paper we explore the dual concepts of maximising geometric uniformity and also local symmetry during refinement of atomistic models of amorphous materials. The first of these corresponds to a more subtle and powerful implementation of the INVERT approach than its previous incarnation, where we now also reduce variance in higher-order correlation functions despite these functions not being directly fixed by the PDF. The second concept---that of maximising local symmetry---reflects our long-term goal of guiding refinement towards the simplest structural solution consistent with the PDF. We demonstrate that the measures of geometric uniformity and local symmetry do indeed act as suitable metrics for the quality of configurations. But, unfortunately, we also find that in their simplest implementations these metrics are unable to drive refinement, implying that the corresponding configurational landscape is too complex. We conclude with some discussion of how we anticipate these ideas may be advanced in future studies.
  
\section{Descriptors of local geometry and symmetry}
\begin{figure}
\includegraphics*[width=\linewidth]{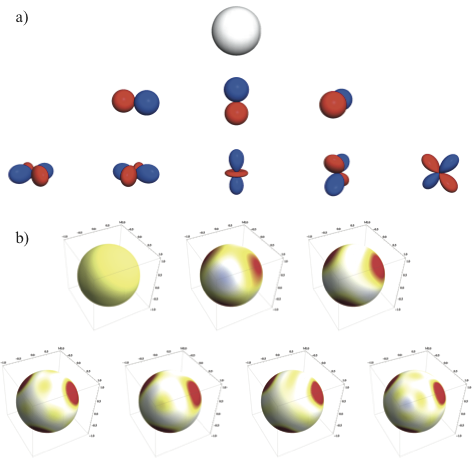}
\caption{%
(a) Graphical representations of the real forms of the spherical harmonics, displayed as $r=Y_{l,m}$, for values of $l$ up to $l=3$ (blue represents positive values and red negative values). (b) A demonstration of the reconstruction of a tetrahedral distribution function $F(\theta,\phi)$ by incorporating $q_{l,m}Y_{l,m}(\theta,\phi)$ terms of increasing order $l$ (up to $l=6$).}
\label{fig2}
\end{figure}

Our starting point is to consider how we might quantitatively compare arbitrary local geometries in an atomistic configuration of a highly disordered materials. We consider geometry to consist of two components: the distance between an atom and its neighbours, and also the angular arrangement of those neighbours. The distance specification is straightforward (indeed it is given by the PDF); this is precisely the metric used in our earlier INVERT study \cite{Cliffe2010}. Angular distributions are most naturally described using the language of spherical harmonics (which are familiar as the angular part of the hydrogenic atomic orbitals [Fig.~\ref{fig2}(a)]). By projecting the orientational distribution of neighbours onto the spherical harmonics, the coefficients so obtained quantify the angular component of this distribution; in this sense we might consider the spherical harmonic coefficients as quantitative ``descriptors'' of the angular distribution of interatomic vectors. Spherical harmonics also provide a straightforward means by which to address the problem of ``registration'' for disordered configurations, by which we mean the absence of a global set of reference axes. By taking the average of the coefficients $q_{l,m}$ for each spherical harmonic $Y_{l,m}$ over all allowed values $m$ for a given index $l$, then that average becomes a quantitative descriptor that is itself independent of any set of reference axes. A similar approach is used in the field of computer science when addressing the general problem of three-dimensional shape matching \cite{Novotni2003,Kazhdan2003}.

Formally, we choose to describe the atomic environment of a given atom in terms of the distribution function $F(\theta,\phi)$, given as the sum of Dirac delta functions positioned at each neighbour position, where in the simplest case we neglect the radial distance:
\begin{equation}
F_i(\theta,\phi)=\sum_{j}\delta(\theta_{ij},\phi_{ij}).\label{Fsum} 
\end{equation}
Projection onto the spherical harmonics yields the $q_{l,m}$:
\begin{equation}
F(\theta,\phi)=\sum_{l=0}^{\infty}\sum_{m=-l}^{l}q_{l,m}Y_{l,m}(\theta,\phi),
\end{equation}
where $Y_{l,m}(\theta,\phi)$ is the spherical harmonic of order $l$ and $m$. The coefficients $q_{l,m}$ can then be averaged across $m$ and normalised to yield our rotationally-invariant descriptors $Q_l$:
\begin{equation}
 Q_l = \sqrt{\frac{4\pi}{(2l+1)} \left( \sum_{m=-l}^{l}q_{l,m} \right)^2}.
\end{equation}

\begin{figure}
\includegraphics*[width=\linewidth]{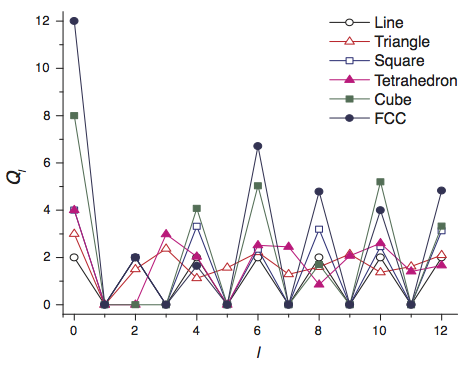}
\caption{%
Values of the orientationally-invariant descriptors $Q_l$ determined for a range of common high-symmetry local environments.}
\label{fig3}
\end{figure}

These $Q_l$ are precisely the quantities used frequently in the fields of amorphous materials and liquids as a means of characterising local geometry  \cite{Steinhardt1983,Baranyai1987,LeRoux2010}. There exist relatively intuitive meanings to the $Q_l$: for example, the $l=0$ term is a simple measure of the number of neighbours, $l=1$ describes the extent to which the local environment is polar, and $l=2$ the extent to which the environment is quadrupolar. The values of $Q_l$ calculated for a variety of common coordination environments are shown in Fig.~\ref{fig3}.

Having established in the $Q_l$ a metric that reflects the geometric arrangement of neighbours around a central atom, we seek now to determine a second metric that represents the degree of symmetry of that same arrangement. Whereas for crystalline materials one has access to the space groups formalism and the group/subgroup relations that provides, an arbitrarily-applicable metric of local symmetry suited to amorphous materials is not so obvious. Certainly at the macroscopic, continuum scale, amorphous materials are isotropic. But the local arrangements of neighbours around any given atom can assume their own approximate point symmetry, and it is this symmetry that we seek to characterise (and ultimately maximise during refinement). We require a number of properties for our metric: first, like the $Q_l$ it too should be orientationally-invariant; second, for two environments with an obvious point group/subgroup relation (\emph{e.g.} tetrahedral and cubic) then this ordering should be robustly reflected in the value of the metric; third, the metric should vary in a continuous way for distortions away from ideal point symmetries in order to allow refinement; and, fourth, the metric must be independent of the particular system of interest. By this last point we are trying to exclude metrics tailored to specific geometries---\emph{e.g.} determining the deviation away from tetrahedral angles---because in such a case one would have to decide \emph{a priori} which particular manifestation of the metric (and hence which geometry) was appropriate for that system.

Inspection of the variation in $Q_l$ for different local symmetries [Fig.~\ref{fig3}] suggests a possibility for one such metric (although there are likely to be others): namely that the variance in $Q_l$ across $l$ is largest for the most symmetric arrangements. Specifically, higher-symmetry arrangements (\emph{e.g.} cubic or FCC) show a higher degree of variability amongst their $Q_l$ values than lower symmetry environments (\emph{e.g.} triangular). Indeed the requisite point group/subgroup relations also emerge: the cubic arrangement (point group $O_h$) has many more zero values of $Q_l$ than does the tetrahedral arrangement (point group $T_d\subset O_h$). A large variance in $Q_l$ for high-symmetry environments is straightforwardly rationalised in terms of the observation that such arrangements correspond to a greater degree of ``interference'' between harmonics, both destructive and constructive. This also suggests that a quantitative measure of symmetry (which we denote $S$) might be given by the variance of $Q_l$ across $l$:
\begin{equation}
S=\frac{1}{l_{\textrm{max}}}\sum_{l=1}^{l_{\textrm{max}}}\left(\frac{Q_l}{\bar Q}-1\right)^2,
\end{equation}
where
\begin{equation}
\bar Q=\sum_{l=1}^{l_{\textrm{max}}}Q_l\\.
\end{equation}

\begin{figure}
\includegraphics*[width=\linewidth]{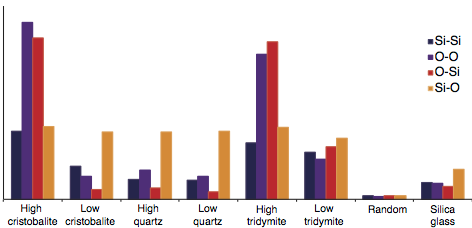}
\caption{%
Values of the symmetry descriptor $S$ for a number of SiO$_{2}$ polymorphs.}
\label{fig4}
\end{figure}

Our next step was to verify whether this new metric $S$ quantitatively reflects our intuitive understanding of symmetry for real materials. We present here the specific results for the the variation in $S$ across a range of tetrahedral SiO$_2$ polymorphs, but similar results are obtained for many different systems. We focus on silica because of its diversity of polymorph symmetries and also its well-understood group/subgroup relations. Because there are two types of atom in SiO$_2$ there are in fact four $S$ terms---one for each possible pair of atom types, counting both orderings for each pair individually (\emph{e.g.} arrangement of O atoms around Si atoms; arrangement of Si atoms around O atoms). In all silica polymorphs the tetrahedral SiO$_4$ geometries are well preserved and so we would expect $S_{\textrm{Si--O}}$ to remain essentially constant;\footnote{We note that we are using configurations based on the average structure for each polymorph, so there are no geometric distortions other than those already present in the crystal structure itself.} however because the various different polymorphs contain different inter-tetrahedral angles we might anticipate that $S_{\textrm{O--O}}$, $S_{\textrm{O--Si}}$ and $S_{\textrm{Si--Si}}$ should vary in a manner that depends on the symmetry of the overall crystal packing. These expectations are indeed borne out in our calculations [Fig.~\ref{fig4}]. In particular, for all crystalline polymorphs the value of $S_{\textrm{Si--O}}$ is essentially invariant, and this value is not very much lower in the model of silica glass reported in Ref.~\cite{Tucker2005}. For each group/subgroup relation (\emph{e.g.} high-quartz/low-quartz), each of the values reduce. Moreover, the high symmetries of the $\beta$-cristobalite and $\beta$-tridymite packings (which are well understood to be anomalously high for good physical reasons \cite{Tucker2001,Dove2000}) are immediately evident from the substantially larger values of $S_{\textrm{O--O}}$, $S_{\textrm{O--Si}}$ and $S_{\textrm{Si--Si}}$. Finally, a random arrangement of Si and O atoms produces very small values of $S$ that essentially represent the uncertainty associated with our calculations.

\begin{figure}
\includegraphics*[width=\linewidth]{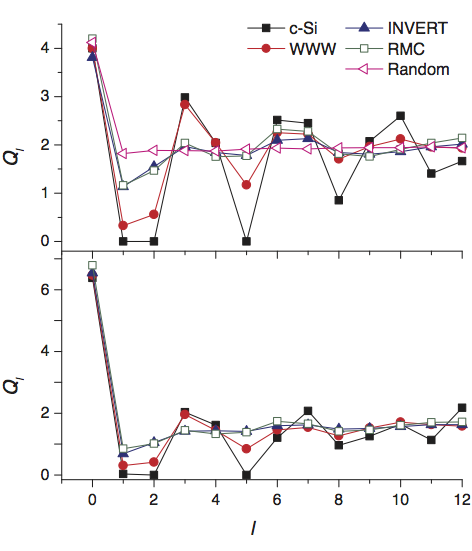}[t]
\caption{%
Values of $Q(l)$ determined for a variety of Si configurations using (top) nearest-neighbours only, with $r_{\textrm{cut}}$ = 2.3\,\AA, (bottom) all neighbours, weighted by a term proportional to $r^{-4}$.}
\label{fig5}
\end{figure}

\section{Case-study: \emph{a}-Si}
Having established suitable metrics both for local geometry and local symmetry, we now proceed to explore how these might be applied to understanding and refining the structure of the canonical disordered system \emph{a}-Si. In our analysis, we make use of five different \emph{a}-Si configurations, each of which contains the same number of atoms and is of the same physical dimensions. The atomic coordinates in these configurations have been chosen or refined in the following ways: (i) randomly, (ii) by fitting this same random configuration to an ideal \emph{a}-Si PDF using RMC, (iii) by fitting to the same PDF using RMC+INVERT, (iv) taken from a high-quality WWW run, and (v) taken from a model of crystalline silicon. For clarity, models (ii), (iii) and (iv) all share essentially the same PDF, whereas model (i) is too random and model (v) is too ordered.

In Fig.~\ref{fig5} we compare the successive values of $Q_l$ for each of these five models. What is immediately clear is that the more ordered configurations correspond to more strongly-varying $Q_l$ series. Importantly, while models (ii)--(iv) share the same PDF there is a substantial difference in the extent of orientational order in models (ii) and (iii) on the one hand, and model (iv) on the other hand. Hence this metric is discriminating for quality of structural model in a way that is not possible using either fit-to-data or local bond length variance as metrics.

\begin{figure}
\includegraphics*[width=\linewidth]{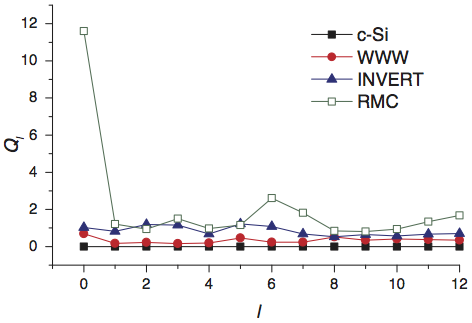}
\caption{%
Values of Var($Q_l$) determined for the set of Si configurations represented in Fig.~\ref{fig5}, making use of an $r^{-4}$ weighting.}
\label{fig6}
\end{figure}

Keeping in mind an eventual aim of using the $Q_l$ in refinement, we explored also the possibility of including in the sum of Eq.~\eqref{Fsum} all atoms in a configuration, with the $\delta$ functions weighted by a term that depends as the inverse fourth power of the interatomic distance. This has the effect of continuing to weight most heavily the local geometry but at the same time allows for a continuous variation in interatomic separations as atoms approach or move away from each other during refinement.\footnote{If Eq.~\eqref{Fsum} is taken as simply the sum over the four nearest neighbours then a cut-off distance $r_{\textrm{cut}}$ must be defined in order to determine which pairs of atoms are nearest neighbours and which are not. The existence of such a cut-off leads to spurious refinement behaviour near $r_{\textrm{cut}}$.} Naturally, other weighting strategies with similar $r$-dependence (\emph{e.g.}\ exponential or Gaussian decay) will be possible; our choice here has been motivated primarily by to ease of calculation and we have not rigorously tested the effect of different weightings on the results obtained. We find that the use of an $r^{-4}$ weighting does not have a large effect on the $Q_l$ variance observed originally [Fig.~\ref{fig5}]. For completeness we note that the value of $Q_0$ increases by $\sim$50\% from its original value of 4 (\emph{i.e.}\ the average coordination number) and hence will dominate the variance metric $S$ more strongly as a result. We have also had to exclude the random model (i) from the corresponding plot in Fig.~\ref{fig5} since the close approach of some pairs of atoms in a randomly-populated configuration leads to unphysical values of the $Q_l$.

Our ultimate refinement strategy involves two new aspects: the first is to \emph{minimise} the variance in the $Q_l$ values amongst all atoms in the configuration as refinement proceeds (to ensure that the local geometries are as similar as possible) and the second is to \emph{maximise} the variance in the total (average) $Q_l$ across different values of $l$ (to ensure that this shared geometry has the highest possible local symmetry). While the data shown in Fig.~\ref{fig5} suggest that the latter metric behaves appropriately for our five configurations, in order to evaluate the former it is necessary to determine the atomic variance in $Q_l$ for these same configurations. These data are given in Fig.~\ref{fig6}. Again the expected trend is observed: specifically, the greater the structural simplicity, the lower the Var($Q_l$) values. We note that Var($Q_0$) corresponds to the variance in coordination number; that this value is so large for the RMC configuration helps explain why RMC+INVERT performs so well at improving the RMC-only coordination number distribution \cite{Cliffe2010}.

\begin{figure}
\includegraphics*[width=\linewidth]{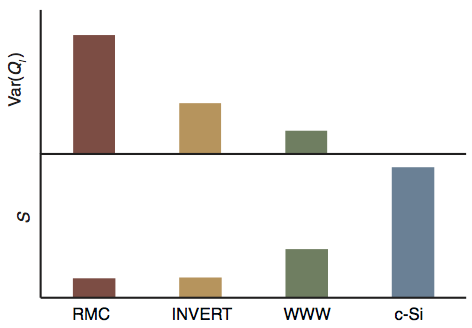}
\caption{%
A plot comparing the values of Var$Q(l)$ and $S$ for a number of amorphous silicon configurations. More ordered configurations consistently take higher values of $S$ and lower values of Var$Q(l)$.}
\label{fig7}
\end{figure}

\begin{figure*}%
\includegraphics*[width=\textwidth]{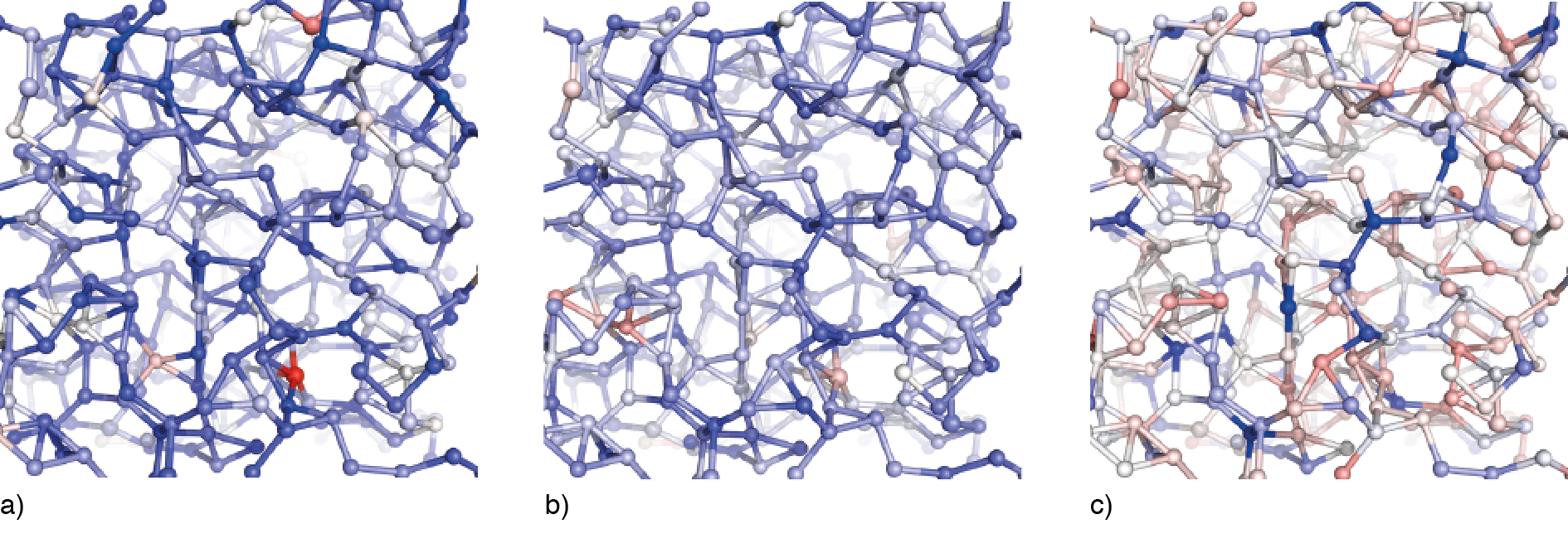}
\caption{A single configuration of \emph{a}-Si produced using the INVERT+RMC of Ref.~\cite{Cliffe2010}, with each panel coloured according to the goodness of fit for each of three metrics (red is worse, blue is better): (a) PDF variance, (b) $Q(l)$ variance, and (c) $S$ symmetry measure.}
\label{fig8}
\end{figure*}

An equivalent comparison can be performed for the symmetry metric $S$ (noting that, in this case, because there is only one atom type there is only one such value for each configuration). The values of $S$ obtained are plotted in Fig.~\ref{fig7}, together with the variance terms obtained by summing the individual Var$(Q_l)$ terms of Fig.~\ref{fig6}. We arrive again at the expected conclusion that $S$ increases as the material becomes more symmetric. Perhaps more surprisingly, the value of $S$ for the RMC+INVERT configuration is very similar to that obtained for the RMC-only refinement, suggesting that the INVERT implementation of Ref.~\cite{Cliffe2010} does little to increase the symmetry of Si arrangements during refinement. We illustrate this in Fig.~\ref{fig8}, where we show part of a RMC+INVERT configuration coloured according to the three metrics we consider: namely, the original INVERT term that characterises variance in interatomic separations, the variance in local $Q_l$ values, and finally the symmetry metric $S$. What we observe is that the three metrics are orthogonal: parts of the configuration that are ranked highly by one are often ranked poorly by another. It is also clear that the symmetry metric $S$ has substantially different character to the two variance constraints. We consider this a promising result, since it suggests that the symmetry metric captures some important aspect of local structure in real CRNs that must surely be maximised in any eventual refinement strategy.

Our final step involved implementation of these local metrics as additional weighting factors within an RMC refinement program, following the general process outlined in Ref.~\cite{Cliffe2010} but noting that in this instance that the metric $S$ is to be \emph{increased} rather than decreased during refinement. Using \emph{a}-Si as a model system, we found that refinements were capable of reducing slightly the value of Var$(Q_l)$ and increasing slightly the value of $S$ relative to the values obtained in a separate RMC+INVERT refinement, but that we were unable to produce PDF fits of sufficient quality at the same time. Since we know that solutions are possible (indeed the WWW model (iv) above is one), what this suggests is that the configurational landscape is too unfavourably structured in this particular implementation to access the set of structurally equivalent global minima. Having established the quality of the individual metrics, this is of course unfortunate that refinement does not appear to be possible using these same metrics. Nevertheless there remain a number of areas where improvement of the refinement process is possible, such as varying the distance weighting regime used or modifying the move selection strategy.

\section{Conclusions}
In the absence of a definitive refinement strategy, perhaps the key result of this paper has been to demonstrate the existence of meaningful metrics the characterise local arrangement of neighbours in configurations of highly disordered materials. Our approach is independent of ``ideal'' geometry such that the methods as developed here should be equally applicable to CRNs of arbitrary local symmetry. We identify also the possibility of a ``parsimony''-driven refinement strategy for disordered materials, whereby solutions that are equally consistent with diffraction data are discriminated on the basis of structural simplicity. Our metrics then provide the mechanism by which that level of simplicity can be determined. Our focus here has been on chemically simple systems, but we anticipate that these same general concepts will extend naturally to more complex systems. One such example is that of plastic phases, in which the constituent molecules have translational order but are orientationally disordered. In such cases one can easily envisage that the orientational distribution functions show strong order over short length scales, and that these distributions are similar from one site to the next despite the inevitable variation in local reference frame. Finally, we suggest that the invariants we develop may also provide a valuable route for configurational analysis irrespective of whether or not they are used to drive refinement itself.

\begin{acknowledgement}
The authors gratefully acknowledge valuable discussions with D.~A.~Drabold and are grateful to the EPSRC and the ERC for financial support under grants EP/G004528/2 and 279705, respectively. The spherical harmonics calculations described here made use of the freely available software archive {\sc shtools} (\emph{shtools.ipgp.fr}), for which we are also grateful.
\end{acknowledgement}

%\bibliographystyle{pss}
%\bibliography{INVERT2}

\begin{thebibliography}{[10]}

\othercit
\bibitem{Egami2003}% book
 \textsc{T.~Egami} and  \textsc{S.\,J.\,L. Billinge},
{Underneath the Bragg Peaks, Structural Analysis of Complex Materials}
  (Pergamon, Oxford, 2003).


\bibitem{Juhas2006}% article
 \textsc{P.~Juh\'{a}s},  \textsc{D.\,M. Cherba},  \textsc{P.\,M. Duxbury},
  \textsc{W.\,F. Punch},  and  \textsc{S.\,J.\,L. Billinge}\iffalse {Ab initio
  determination of solid-state nanostructure.}\fi,
 \jr{Nature} \textbf{440}(7084), 655--8 (2006).


\bibitem{Cliffe2010}% article
 \textsc{M.\,J. Cliffe},  \textsc{M.\,T. Dove},  \textsc{D.\,A. Drabold},  and
  \textsc{A.\,L. Goodwin}\iffalse {Structure Determination of Disordered
  Materials from Diffraction Data}\fi,
 \jr{Physical Review Letters} \textbf{104}(12), 1--4 (2010).


\bibitem{Juhas2010}% article
 \textsc{P.~Juh\'{a}s},  \textsc{L.~Granlund},  \textsc{S.\,R. Gujarathi},
  \textsc{P.\,M. Duxbury},  and  \textsc{S.\,J.\,L. Billinge}\iffalse {Crystal
  structure solution from experimentally determined atomic pair distribution
  functions}\fi,
 \jr{Journal of Applied Crystallography} \textbf{43}(3), 623--629 (2010).


\bibitem{Weiner2005}% article
 \textsc{S.~Weiner},  \textsc{I.~Sagi},  and  \textsc{L.~Addadi}\iffalse
  {Structural biology. Choosing the crystallization path less traveled.}\fi,
 \jr{Science (New York, N.Y.)} \textbf{309}(5737), 1027--8 (2005).


\bibitem{Michel2007}% article
 \textsc{F.\,M. Michel},  \textsc{L.~Ehm},  \textsc{S.\,M. Antao},
  \textsc{P.\,L. Lee},  \textsc{P.\,J. Chupas},  \textsc{G.~Liu},
  \textsc{D.\,R. Strongin},  \textsc{M.\,a.\,a. Schoonen},  \textsc{B.\,L.
  Phillips},  and  \textsc{J.\,B. Parise}\iffalse {The structure of
  ferrihydrite, a nanocrystalline material.}\fi,
 \jr{Science (New York, N.Y.)} \textbf{316}(5832), 1726--9 (2007).


\bibitem{Walsh2009}% article
 \textsc{A.~Walsh},  \textsc{J.\,L.\,F. {Da Silva}},  and  \textsc{S.\,H.
  Wei}\iffalse {Interplay between Order and Disorder in the High Performance of
  Amorphous Transparent Conducting Oxides}\fi,
 \jr{Chemistry of Materials} \textbf{21}(21), 5119--5124 (2009).


\bibitem{Sun2006}% article
 \textsc{Z.~Sun},  \textsc{J.~Zhou},  and  \textsc{R.~Ahuja}\iffalse {Structure
  of Phase Change Materials for Data Storage}\fi,
 \jr{Physical Review Letters} \textbf{96}(5), 055507 (2006).

\bibitem{Welberry1994}% article
 \textsc{T.\,R. Welberry} and  \textsc{B.\,D. Butler}\iffalse {Interpretation
  of diffuse X-ray scattering via models of disorder}\fi,
 \jr{Journal of Applied Crystallography} \textbf{27}(3), 205--231 (1994).

\bibitem{Evans1990}% article
 \textsc{R.~Evans}\iffalse {Comment on Reverse Monte Carlo Simulation}\fi,
 \jr{Molecular Simulation} \textbf{4}(6), 409--411 (1990).

\bibitem{Wooten1987}% article
 \textsc{F.~Wooten} and  \textsc{D.~Weaire}\iffalse {Modeling Tetrahedrally
  Bonded Random Networks by Computer}\fi,
 \jr{Solid State Physics} \textbf{40}(null), 1--42 (1987).


\bibitem{Wooten1985}% article
 \textsc{F.~Wooten},  \textsc{K.~Winer},  and  \textsc{D.~Weaire}\iffalse
  {Computer Generation of Structural Models of Amorphous Si and Ge}\fi,
 \jr{Physical Review Letters} \textbf{54}(13), 1392--1395 (1985).



\bibitem{Biswas2004}% article
 \textsc{P.~Biswas},  \textsc{R.~Atta-Fynn},  and  \textsc{D.~Drabold}\iffalse
  {Reverse Monte Carlo modeling of amorphous silicon}\fi,
 \jr{Physical Review B} \textbf{69}(19), 195207 (2004).


\bibitem{Muller-Warmuth1982}% article
 \textsc{W.~M\"{u}ller-Warmuth} and  \textsc{H.~Eckert}\iffalse {Nuclear
  magnetic resonance and M\"{o}ssbauer spectroscopy of glasses}\fi,
 \jr{Physics Reports} \textbf{88}(2), 91--149 (1982).


\bibitem{Brodsky1978}% article
 \textsc{M.~Brodsky} and  \textsc{M.~Cardona}\iffalse {Local order as
  determined by electronic and vibrational spectroscopy: Amorphous
  semiconductors}\fi,
 \jr{Journal of Non-Crystalline Solids} \textbf{31}(1-2), 81--108 (1978).


\bibitem{Gereben1994}% article
 \textsc{O.~Gereben} and  \textsc{L.~Pusztai}\iffalse {Structure of amorphous
  semiconductors: Reverse Monte Carlo studies on a-C, a-Si, and a-Ge}\fi,
 \jr{Physical Review B} \textbf{50}(19), 14136--14143 (1994).


\bibitem{Pauling1929}% article
 \textsc{L.~Pauling}\iffalse {THE PRINCIPLES DETERMINING THE STRUCTURE OF
  COMPLEX IONIC CRYSTALS}\fi,
 \jr{Journal of the American Chemical Society} \textbf{51}(4), 1010--1026 (1929).
 
 \bibitem{Shao1990}
  \textsc{ W.-L.~Shao},   \textsc{J.~Shinar},  \textsc{B. C.~Gerstein},   \textsc{F.~Li}, and   \textsc{J. S.~Lannin},
 \jr{ Physical Review B} \textbf{41}(13), 9491-Ð9494 (1990).

\othercit
\bibitem{Novotni2003}% inproceedings
 \textsc{M.~Novotni} and  \textsc{R.~Klein},
{3D zernike descriptors for content based shape retrieval},
 in: Proceedings of the eighth ACM symposium on Solid modeling and applications
  - SM '03,  (ACM Press, New York, New York, USA, June 2003),  p.\,216.


\othercit
\bibitem{Kazhdan2003}% inproceedings
 \textsc{M.~Kazhdan},  \textsc{T.~Funkhouser},  and  \textsc{S.~Rusinkiewicz},
{Rotation Invariant Spherical Harmonic Representation of 3D Shape Descriptors},
 in: Eurographics Symposium on Geometry Processing,  (2003).


\bibitem{Steinhardt1983}% article
 \textsc{P.~Steinhardt},  \textsc{D.~Nelson},  and
  \textsc{M.~Ronchetti}\iffalse {Bond-orientational order in liquids and
  glasses}\fi,
 \jr{Physical Review B} \textbf{28}(2), 784--805 (1983).


\bibitem{Baranyai1987}% article
 \textsc{A.~Baranyai},  \textsc{A.~Geiger},  \textsc{P.\,R. Gartrell-Mills},
  \textsc{K.~Heinzinger},  \textsc{R.~McGreevy},  \textsc{G.~P�link�s},
  and  \textsc{I.~Ruff}\iffalse {Invariants of spherical harmonics as order
  parameters in liquids}\fi,
 \jr{Journal of the Chemical Society, Faraday Transactions 2} \textbf{83}(8),
  1335 (1987).


\bibitem{LeRoux2010}% article
 \textsc{S.~{Le Roux}} and  \textsc{V.~Petkov}\iffalse {ISAACS – interactive
  structure analysis of amorphous and crystalline systems}\fi,
 \jr{Journal of Applied Crystallography} \textbf{43}(1), 181--185 (2010).


\bibitem{Tucker2005}% article
 \textsc{M.\,G. Tucker},  \textsc{D.\,a. Keen},  \textsc{M.\,T. Dove},  and
  \textsc{K.~Trachenko}\iffalse {Refinement of the Si–O–Si bond angle
  distribution in vitreous silica}\fi,
 \jr{Journal of Physics: Condensed Matter} \textbf{17}(5), S67--S75 (2005).


\bibitem{Tucker2001}% article
 \textsc{M.\,G. Tucker},  \textsc{M.\,P. Squires},  \textsc{M.\,T. Dove},  and
  \textsc{D.\,a. Keen}\iffalse {Dynamic structural disorder in cristobalite:
  neutron total scattering measurement and reverse Monte Carlo modelling}\fi,
 \jr{Journal of Physics: Condensed Matter} \textbf{13}(3), 403--423 (2001).


\bibitem{Dove2000}% article
 \textsc{M.\,T. Dove},  \textsc{A.\,K.\,A. Pryde},  and  \textsc{D.\,A.
  Keen}\iffalse {Phase transitions in tridymite studied using ` Rigid Unit Mode
  ' theory , Reverse Monte Carlo methods and molecular dynamics
  simulations}\fi,
 \jr{Mineralogical Magazine} \textbf{64}, 267--283 (2000).


\end{thebibliography}

\providecommand{\WileyBibTextsc}{}
\let\textsc\WileyBibTextsc
\providecommand{\othercit}{}
\providecommand{\jr}[1]{#1}
\providecommand{\etal}{~et~al.}

\end{document}